# Aberration of starlight experiment

Robert A. Woodruff

Center for Astrophysics and Space Astronomy, Astrophysics Research Laboratory, University of Colorado, 593 UCB, Boulder, Colorado 80309–0593, USA (raw@Colorado.edu)

## ABSTRACT

We propose an experiment using a conventional optical telescope to determine whether aberration of starlight results from special relativistic effects external to a measurement sensor or from optical effects within a sensor. The proposed measurements would discriminate between the two starlight aberration models in an Earth-based experiment. In addition, the measurements would yield an independent experimental test of relativistic time dilation.

## 1. INTRODUCTION

In 1728, Bradley [1] detected aberration of starlight, an annual variation of the apparent position of stars on the celestial sphere, using a simple Earth-based telescope. The currently accepted theory for starlight aberration flows from special relativity and is presented in numerous texts, for instance French [2]. Woodruff [3] applied physical optics imaging theory to obtain an alternative theoretical model for aberration of starlight. He proposed an optical sensor that uses a solid glass telescope to distinguish between the two models.

The special relativistic-based model attributes stellar aberration to tilt of the incident plane wavefronts. Presumably the wavefront tilts external to the measurement sensor and results from time dilation in the sensor frame. [4] The physical optics-based model considers untilted converging wavefronts within the sensor to explain aberration of starlight phenomenon. In this model, the wavefronts shear laterally in the sensor frame, as they converge to form an image of the star, due to the translation of the sensor relative to the converging wavefronts. Predictions of the two models agree for relatively slow sensor motions, for instance at Earth orbital velocity, but differ greatly at velocities approaching the speed of light. This paper presents another optical experiment configured specifically to test aberration at Earth orbital velocity relative to these two models.

## 2. TWO THEORIES FOR ABERRATION OF STARLIGHT

### A. Adopted geometry

Following Woodruff [3], Figure 1 presents the overall geometry of a typical aberration measurement using a simple telescope. A distant star emits light into free space as spherical wavefronts. These expand in all directions as they propagate through space and over a great distance, where they reach the observer, effectively becoming planar of infinite extent. Figure 1 defines a Cartesian coordinate system (x, y, z) oriented so the star's unaberrated position lies along the positive y-axis. The incident Poynting vector, which is normal to the incident planar wavefronts, is parallel to the y-axis and therefore parallel to the x-z plane. The aberrated position of the star is on a line in object space, labeled as the y′-axis, between the center of the sensor aperture and the measured location of the star on the sky. When the sensor optical axis oriented parallel to the y-axis, the acute angle in the x-y plane between the y′-axis and the optical axis is the aberration angle $\alpha(\theta)$. The velocity vector $\vec{v}$ of the sensor lies in the x-y plane at an angle $\theta$ relative to the +y-axis and therefore at an angle $(\pi/2) - \theta$ relative to the incident plane wavefronts. The y′-axis lies in the x-y plane. Resolved into this plane, the magnitude of component of the sensor velocity parallel to the incident wavefront is $v_x = v \sin(\theta)$ and the magnitude of the component perpendicular to the incident plane wavefront is $v_y = v \cos(\theta)$.

### B. Review of two aberration of starlight theories

In the physical optics-based model [3], the aberration angle for a vacuum-filled sensor satisfies

$$\alpha(\theta) = \arctan(v_x / c) = \arctan\left\{\frac{v \sin(\theta)}{c}\right\}, \text{ where } c \text{ is the speed of light in vacuum.} \qquad (1)$$

Stellar aberration is zero when $\theta = 0$. The maximum aberration geometry results when the sensor motion is parallel to the incident plane wavefronts from the star, i.e., when $\theta = \pi/2$. The aberration angle is then given by

$$\alpha(\pi/2) = \arctan\{\beta\} \text{ where } \beta = v/c. \qquad (2)$$

At Earth-orbital velocities, where $\beta \ll 1$, the maximum value is

$$\alpha(\pi/2) \approx \beta \text{ if } \beta \ll 1. \qquad (3)$$

An Earth-based experiment would conclude that the aberration angle varies from zero at $\theta = 0$ to a maximum value of $\alpha(\pi/2) \approx \beta \approx 10^{-4}$ [5] matching Bradley's result. Note that Equation (2) predicts

$$\alpha(\frac{\pi}{2}) = 45 \text{ degrees if the sensor velocity were the speed of light } v = c.$$

This model ascribes the effect as due to uniform sensor translation in the plane parallel to the incident plane wavefront during the temporal interval from the instant that a given wavefront undergoes the image forming phase shift by the imaging optic to the instance that the resultant converging wavefront reaches the detector forming an image of the star. In the sensor's reference frame, the image forms at coordinates $(-f_0 \tan(\alpha), -f_0, 0)$ where $f_0$ is the paraxial focal length of the optic. Thus, the star appears off-axis to the sensor by the angle given in Equation (1). In the sensor frame, the converging wavefronts appear to translate perpendicular to the optical axis as they converge to form the stellar image. The converging spherical wavefronts appear to shear, but do not tilt.

The Special Theory of Relativity attributes starlight aberration to time dilation across the sensor aperture causing the incident plane wavefronts to appear tilted [4]. It finds that $\alpha(\theta)$ is the solution to [2]

$$\cos(\theta - \alpha(\theta)) = [\cos(\theta) + \beta] / [1 + \beta \cos(\theta)] \qquad (4)$$

Based on Equation (4) the aberration angle is zero when $\theta = 0$, as expected. The maximum aberration angle occurs when $\theta = \pi/2$ with the maximum aberration given by

$$\alpha(\pi/2) = \arcsin\{\beta\}. \tag{5}$$

For an Earth-based observation where $\beta << 1$, the maximum aberration angle is therefore

$$\alpha(\pi/2) \approx \beta \text{ if } \beta << 1. \tag{6}$$

Equation (3) is identical to Equation (6). The two models are indistinguishable at small velocities. However, based on Equations (1) and (4), predictions of the two models differ greatly at larger sensor velocities. No differences occur at any velocity for $\theta = 0$. For $\theta = \pi/2$ the differences are small even for relatively large sensor velocity, e.g., the measurements would differ by only 12.9 milli-arc-seconds when $\beta = 0.005$, i.e. approximately 50 times Earth orbital velocity. The difference is only 2.8 arc-seconds at $\beta = 0.03$, approximately 300 times Earth orbit velocity. However, at the speed of light, a significant difference occurs with the relativistic-based theory predicting $\alpha(\frac{\pi}{2}) = 90^o$ and the wave-based theory predicting $\alpha(\frac{\pi}{2}) = 45^o$.

## 3. NEW EXPERIMENTAL TEST

Based on Equations (2) and (5), the difference predicted for the maximum value of aberration by the two models in Earth orbit velocity experiments is extremely small, approximately $10^{-7}$ arc-seconds. Experiments with simple single air-filled telescopes at Earth orbital velocity do not possibly have sufficient measurement precision to discern between the two models. However, by exploiting the physical characteristics incorporated within the derivation of the models, a slightly more intricate optical configuration can be configured that would break this degeneracy and discriminate between the models. Each of the two models incorporates distinct experimentally testable physical characteristics. The special relativistic-based model ascribes the aberration effect to sensor motion relative to the incident wavefront independent to the characteristics of the sensor. [6] It models aberration as the result of wavefront tilt occurring external to and independent of the sensor geometry. Thus in this model, aberration should be independent of telescope optical properties. The physical optics-based model predicts the measurement will depend on the sensor configuration including upon its internal optical properties.

### A. Instrumentation description

Figure 2 shows a sensor designed to experimentally discern between the two aberration models. The sensor consists of a single optical telescope consisting of one shared imaging optic, one shared optical beamsplitter, a shared single rigid structure, and two identical detectors. One detector, the direct path detector, views the target star with light transmitted by the optical beamsplitter. The folded path detector concurrently views the same star with light folded at 90 degrees by reflection from the first surface of the beamsplitter. All direct and folded optical axes lie in the p-q plane and the viewing direction of the telescope is the +q-axis. In the sensor frame, the axial distance from the imaging optic to the beamsplitter along the direct path optical q-axis is $s_0$, where the subscript "0" refers to the rest frame of the sensor. In a frame moving at relative velocity $v$ parallel to the q-axis, Lorentz contracted distance becomes $s = \frac{s_0}{\gamma}$ with $\gamma = (1 - \beta^2)^{-0.5}$. The rest frame value for the paraxial focal length of the imaging optic is given by $f_0$ and the Lorentz contracted value by $f = \frac{f_0}{\gamma}$.

This simple sensor can be applied to numerous aberration experiments. One experimental geometry, which we will refer to as "Geometry 1", would orient the Earth-orbit velocity parallel to the +q-axis (i.e., $\theta = 0$) with the sensor input optical axis moving directly toward the star. A second experimental geometry, "Geometry 2", would orient the Earth-orbit velocity parallel to the +p-axis, at $\theta = \pi/2$, in the plane of the sensor optical axes (i.e., in the p-q plane). A third experimental geometry, "Geometry 3", would orient the Earth-orbit velocity parallel to the +r-axis, again at $\theta = \pi/2$, but perpendicular to in the plane of the sensor optical axes (i.e., in the q-r plane).

### B. Predictions

We now predict results of measurements by this sensor under the constraints of each of the two models.

*(a) Prediction of special relativistic-based model*

If the special relativistic-based model properly models aberration, both detectors sense zero aberration in the $\theta = 0$ geometry of Geometry 1. In this experimental geometry, an incident plane wavefront encounters the entire optical aperture of the telescope simultaneously and therefore both detectors view an untilted wavefront. Both would record zero aberration.

The $\theta = \pi/2$ geometry would yield identical amounts of aberration for both detectors because time dilation across the laterally moving aperture would result in wavefront tilt at the telescope aperture affecting both optical paths identically independent of sensor configuration. The source would appear off-axis by the same amount in both paths. For Earth-based sensor, the angle would correspond to Bradley's result, $\alpha(\pi/2) \approx \beta$. This conclusion applies for both velocity direction cases, Geometrys 2 and 3, with the velocity in the plane of the sensor optical axes or normal to this plane.

*(b) Prediction of optical physics-based model*

If instead the optical physics-based model properly models aberration, the experimental results would differ from those predicted by the special relativistic-based model. [7]

In the $\theta = 0$ geometry of Geometry 1, the direct path detector senses zero aberration since the spherical wavefronts converge along the direct path optical axis, the q-axis, as it translates parallel to the motion of the sensor. However, the folded detector measures non-zero aberration. In the sensor path along the q-axis between the imaging optic and the beamsplitter, the converging wavefront propagates parallel to the motion of the sensor until it reflects from the beamsplitter. The folded wavefront then converges along a line parallel to the p-axis that is perpendicular to the sensor motion. The converging wavefront will arrive at the beamsplitter at time $t_1 = \dfrac{s}{(c+v)}$ after the instant that the imaging optic introduces a convergent phase shift into the incident plane wave. The beamsplitter folds the center ray of the converging wavefront at this axial location so it propagates parallel to the p-axis. It then images at coordinate $f - s$, in the negative p-direction, at time $t_2 = \dfrac{f - s}{c}$ later than $t_1$. In that time interval the folded path detector translates in the +q-direction parallel by a distance $vt_2 = \beta(f - s)$. The star appears off-axis in the plane of the detector by distance $vt_2$. Therefore, the physical optics-based model predicts that the measured aberration angle for the folded path detector for an experiment in the $\theta = 0$ geometry is the solution to

$$\tan(\alpha) = \frac{vt_2}{f} = \beta\frac{(f - s)}{f} = \beta\frac{(f_0 - s_0)}{f_0} \qquad (7)$$

The result is dependent on the configuration of the sensor, ranging from $\alpha = \arctan(\beta)$ if $s_0 = 0$ to $\alpha = 0$ for $s_0 = f_0$. When $\beta \ll 1$, $\alpha \approx \beta\dfrac{(f_0 - s_0)}{f_0}$. Therefore the folded path detector of an Earth-based sensor in $\theta = 0$ geometry would measure aberration of magnitude between zero and Bradley's result, $\alpha \approx \beta$, depending on the designed location of the beamsplitter within the sensor. The direct path detector would record $\alpha = 0$ independent of the sensor design configuration.

Predictions of optical physics-based model for the sensor used in the $\theta = \pi/2$ geometry of Geometry 2 with velocity in the +p-direction show interesting differences. The direct path detector senses aberration of magnitude given in Equation (1) independent of the sensor design configuration. The folded path detector also measures non-zero aberration. Its converging wavefront propagates from the imaging optic perpendicular to the motion of the sensor until it reflects from the beamsplitter. The folded beam then converges along a line parallel to the p-axis propagating anti-parallel to the sensor motion as it converges to form an image of the star. Again assume the imaging optic introduces a convergent phase shift into the incident plane wave at time $t = 0$. If the beamsplitter is located an axial distance $s_0$ from the imaging optic toward the detector, the converging spherical wavefront will then reach the beamsplitter at time $t_1 = \dfrac{s_0}{(c+v)}$ later than the instant that the imaging optic introduces a convergent phase shift into the incident plane wave. The beamsplitter reflects the wavefront at 90 degrees. The wavefront images at p-coordinate of magnitude $f - s - v(t_1 + t_2)$, in the negative p-direction relative to the q-axis, at time $t_2 = \dfrac{\beta t_1(\gamma - 1)}{(\gamma + \beta)}$ later than $t_1$. During this time interval the sensor with its detector translates a distance $vt_2$ in the +p-direction parallel to the folded optical axis. The star appears off-axis by $vt_1$ viewed in the plane of the detector. The sensor's effective paraxial focal length is of the sum of the distance from the virtual position of the imaging optic, as folded by the beamsplitter, to the beamsplitter, $s_0$, and the Lorentz contracted distance from the beamsplitter to the detector plane $(f - s)$. Therefore the effective paraxial focal length is $\left[ f + s_0\left(1 - \frac{1}{\gamma}\right)\right]$. Therefore, the physical optics-based model predicts that the measured aberration angle for the folded path detector for an experiment in the $\theta = \pi/2$ geometry with sensor velocity in the +p-direction is the solution to

$$\tan(\alpha) = \frac{vt_1}{\left(f + s_0\left(1 - \frac{1}{\gamma}\right)\right)} = \frac{\beta \gamma s_0}{(1-\beta)[f_0 + s_0(\gamma - 1)]} \tag{8}$$

The result is dependent on the configuration of the sensor, ranging from $\alpha = 0$ if $s_0 = 0$ to $\alpha = \arctan\left[\frac{\beta}{(1-\beta)}\right]$ for $s_0 = f_0$. When $\beta \ll 1$, $\alpha \approx \beta \frac{s_0}{f_0}$ for Geometry 2. Therefore the folded path detector of an Earth-based sensor used in the $\theta = \pi/2$ experimental geometry with sensor velocity in the +p-direction would measure aberration of magnitude between zero and Bradley's result, $\alpha \approx \beta$, depending on the location of the beamsplitter within the sensor and the direct path detector would measure $\alpha \approx \beta$ independent of the sensor design configuration.

The optical physics-based model predictions for the sensor used in the $\theta = \pi/2$ geometry with velocity in the +r-direction is easily derived. Both paths move identically in an exact transverse direction relative to the converging wavefronts so both detector sense aberration of magnitude given in Equation (1) independent of the sensor design configuration.

For an Earth-based sensor, the maximum aberration angle is given by $\alpha_E = v_E/c$, i.e., Bradley's result, where $v_E$ is the mean tangential orbital velocity of the Earth relative to the Sun and $c$ is the speed of light in air (or effectively in vacuum). [5] Table 1 summarizes the predicted aberration angles for the proposed experiment using the sensor in Figure 2 with the beamsplitter is placed a fractional distance $\frac{s_0}{f_0}$ from the imaging optic to the detector.

6                                                                                                                                 raw, 6/19/13

| Experimental geometry | Model | Direct path | Folded path |
|---|---|---|---|
| $\theta = 0$<br>Experiment 1 | Special relativistic-based | 0 | 0 |
| | Optical physics-based | 0 | $\alpha \approx \alpha_E \frac{(f_0 - s_0)}{f_0}$ |
| $\theta = \pi/2$, $v$ in p-direction<br>Experiment 2 | Special relativistic-based | $\alpha_E$ | $\alpha_E$ |
| | Optical physics-based | $\alpha_E$ | $\alpha \approx \alpha_E \frac{s_0}{f_0}$ |
| $\theta = \pi/2$, $v$ in r-direction<br>Experiment 3 | Special relativistic-based | $\alpha_E$ | $\alpha_E$ |
| | Optical physics-based | $\alpha_E$ | $\alpha_E$ |

Table 1: Predicted aberration angle measurements for the proposed experiments.

An unambiguous test for the aberration models would result if the sensor viewed Polaris (α UMi Aa) (Right ascension 02$^h$ 31$^m$ 49.09$^s$, declination +89° 15′ 50.8″) sequencing between the Geometry 2 and Geometry 3 geometries. Polaris's celestial location easily allows the $\theta = \pi/2$ geometry from any Northern hemisphere location on the Earth's surface. The +q-axis of the sensor would be fixed on Polaris while we rotate the experiment about the q-axis. The sensor orientation achieves the geometry of Geometry 3 by positioning the p-q plane perpendicular to Earth-orbit motion direction. By rotating the sensor about the q-axis by 90 degrees, we reposition the sensor into the geometry of Geometry 2 with the p-q plane parallel to Earth-orbit motion direction. Table 1 predicts that for both models and both experimental configurations the direct path detector would record full aberration of about 20.5 arc sec. We proceed with the experiment by accepting this result as we perform the measurement, which allows us to rotate the sensor about the direct path star image without introducing result ambiguities as we compare the models. We therefore monitor the position of the star image at the folded path detector. If it does not change between the two geometries, the special relativistic-based model is not excluded and the optical physics-based model is excluded. If it does change by the predicted amount of the physical optics-based model in magnitude and direction, the special relativistic-based model is excluded and the optical physics-based model is not excluded. The optical physics-based model would predict that the aberration angle changes to about 2 arc seconds so the image would move about 18 arc seconds, which is a readily measurable amount of 87 microns with a 1000 mm focal length sensor.

The experiments would clearly distinguish between the two theoretical models.

## 4. CONCLUSION

We presented an experimental optical sensor specifically configured to determine if aberration of starlight phenomena is a manifestation of wavefront tilt external to any measurement sensor or is the result of optical imaging physics within the measurement telescopic sensor. The proposed sensor configuration allows experimental discrimination between wavefront tilt and wavefront shear effects, even at readily accessible slow speeds such as the Earth annual orbital speed. The configuration would experimentally address the question posed by Woodruff [3]: "Does stellar aberration result from wavefront tilt?" The proposed experiment would also provide an independent test of relativistic time dilation, because the relativistic-derived model requires time dilation to induce wavefront tilt external from the sensor.

Acknowledgements

I am grateful to D. Y. Gezari and R. G. Lyon, both of Goddard Space Flight Center, Greenbelt, Maryland, for many inspiring discussions and extremely valuable criticisms, suggestions, and encouragement during formulation and preparation of this paper. I am particularly grateful to anonymous reviewers for much constructive criticism and many useful comments during the peer review of this paper

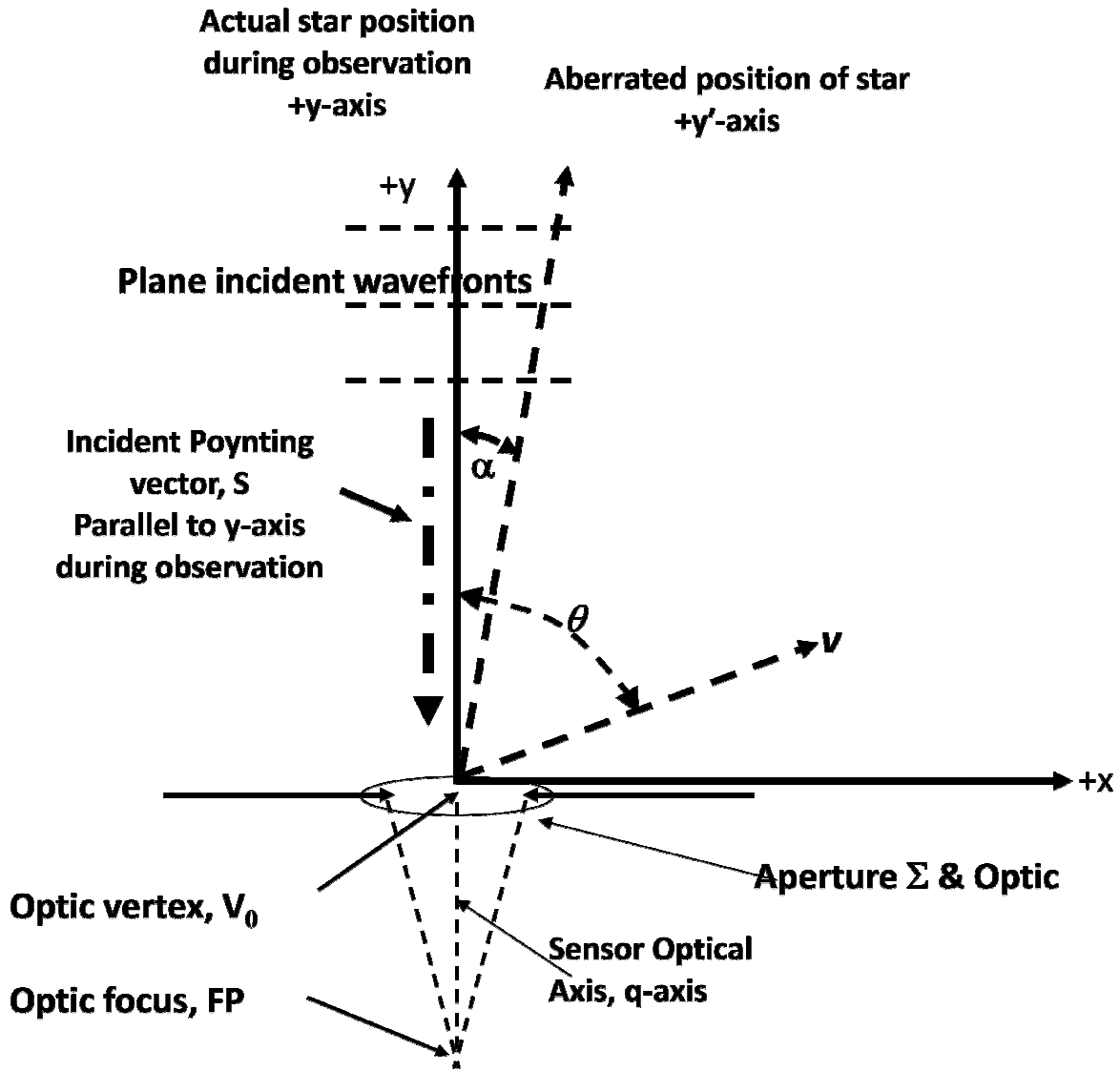

**Figure 1:** Overall geometry
Observational geometry defined in (x, y, z) Cartesian coordinate system.
Sensor geometry defined in (p, q, r) Cartesian coordinate system. Sensor Optical Axis (q-axis) parallel to line of sight to star during observation. Incident plane wavefronts from star lie perpendicular to y-axis. Poynting vector, $\vec{S}$, parallel to y-axis, and positive in –y direction. Angle, $\theta$, measured in x-y plane, between y-axis and sensor velocity direction $\vec{v}$. The aberrated location of star lies along $+y'$ axis in x-y plane. Aberration angle is $\alpha$ in x-y plane. Shape Σ defines optics aperture. Optic images star in sensor focal plane. Sensor velocity component parallel to incident wavefronts is $v_x = v\sin(\theta)$.

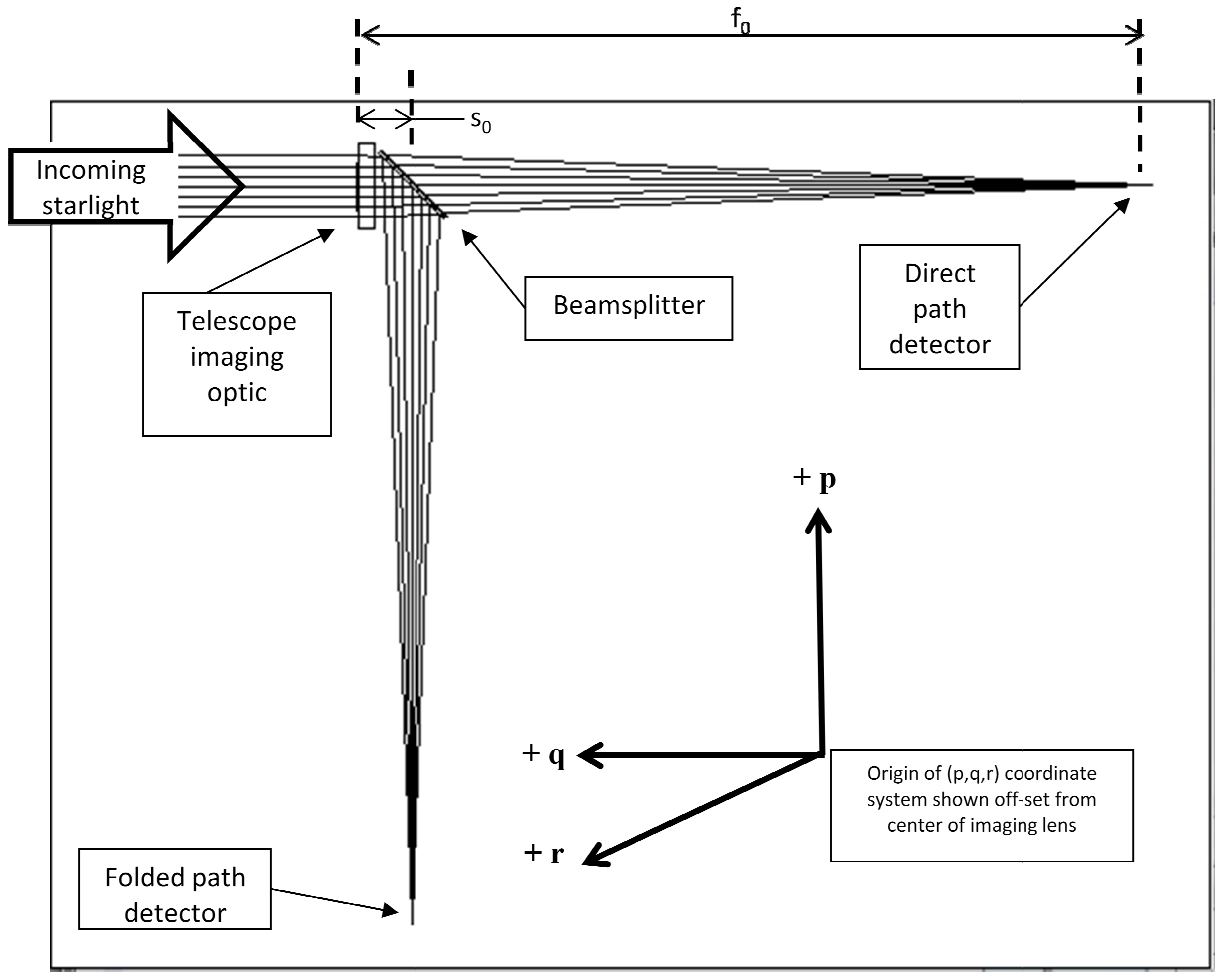

**Figure 2:** Geometry of proposed sensor with two detectors sharing one lens
A single telescope with two identical detectors, mounted on a common rigid structure, views a star. The direct path directly views the imaged star with light transmitted through the optical beamsplitter. The folded path detector simultaneously views the star with light reflected at 90 degrees by the beam splitter's first surface. The beamsplitter lies at rest frame distance $s_0$ along the optical axis from the lens. The rest frame value of the lens paraxial focal length is $f_0$. All optical axes lie in the p-q plane.

In the rest frame the key components lie at the following coordinates (p, q, r):
Imaging optical lens $(0, 0, 0)$
Chief ray intercept at front surface of beamsplitter $(0, -s_0, 0)$
Center of direct path detector $(0, -f_0, 0)$ {ignoring thickness of beamsplitter}
Center of folded path detector $(-[f_0 - s_0], -s_0, 0)$